\documentclass[x11names,12pt]{article}
\usepackage[paper=letterpaper,margin=.85in]{geometry}

\pdfoutput=1

\usepackage{graphicx}				
\usepackage{amsmath}
\usepackage{mathtools}
\usepackage{amssymb}
\usepackage{booktabs}
\usepackage{comment}
\usepackage{amsthm}
\numberwithin{equation}{section}

\usepackage{multirow}
\usepackage{overpic}
\usepackage{caption}
\usepackage{subcaption}
\usepackage{array}
\usepackage{rotating}

\usepackage[utf8]{inputenc}
\usepackage{lmodern}
\usepackage[T1]{fontenc} 
\usepackage{microtype} 

\usepackage{xcolor}

\definecolor{dark-green}{rgb}{0.1,0.4,0}
\definecolor{NiceBlue}{rgb}{0.30196,0.55294,0.57647}

\usepackage{cite}
\usepackage{hyperref}
\hypersetup{
      colorlinks=true,
      linkcolor=dark-green,
      citecolor=Red4,
      urlcolor=dark-green,
      linktoc=page
}
\newcommand{\bea}{\begin{eqnarray}}
\newcommand{\eea}{\end{eqnarray}}

\newcommand\afterTocSpace{\bigskip\medskip}
\newcommand\afterTocRuleSpace{\bigskip\bigskip}

\newcommand{\dd}{\mathrm{d}}

\theoremstyle{definition}

\begin{document} 
\begin{flushright}
  \footnotesize UUITP-17/25\\
  \normalsize
  \end{flushright}
\thispagestyle{empty}

\vspace*{1.5cm}
\begin{center}

{\bf {\LARGE Black Shell Thermodynamics
}}

\begin{center}

\vspace{1cm}

\hypersetup{urlcolor=black}
{\bf \href{mailto:ulf.danielsson@physics.uu.se}{Ulf Danielsson}}$^{1}$, {\bf \href{mailto:vyshnav.vijay.mohan@gmail.com}{Vyshnav Mohan}}$^{1,2}$, \textbf{and} {\bf \href{mailto:lth@hi.is}{L\'arus Thorlacius}}$^{1,2}$

\bigskip

$^{1}$\,Institutionen f\"or fysik och astronomi, Uppsala Universitet\\
Box 803, SE-751 08 Uppsala, Sweden

\bigskip

$^{2}$\,Science Institute, University of Iceland\\
Dunhaga 3, 107 Reykjav\'ik, Iceland

  \end{center}

\vspace{1.5cm}
{\bf Abstract}
\end{center}
\begin{quotation}
\noindent

Black shells have been proposed as black hole mimickers, {\it i.e.} horizonless ultra-compact objects that replace black holes. In this paper, we assume the existence of black shells and consider their thermodynamic properties, but remain agnostic about their wider role in gravitational physics. 
An ambient negative cosmological constant is introduced in order to have a well-defined canonical ensemble, leading to a rich phase structure. In particular, the Hawking-Page transition between thermal AdS vacuum and large AdS black holes is split in two, with an intermediate black shell phase, which may play a role in gauge/gravity duality at finite volume. Similarly, for non-vanishing electric charge below a critical value, a black shell phase separates two black hole phases at low and high temperatures. Above the critical charge, there are no phase transitions and large AdS black holes always have the lowest free energy.

\end{quotation}

\setcounter{page}{0}
\setcounter{tocdepth}{2}
\setcounter{footnote}{0}
\newpage

\parskip 0.1in
 
\setcounter{page}{2}

\setcounter{tocdepth}{1}

{\hypersetup{linkcolor=black}
\tableofcontents
}

\section{Introduction}

Various ultra-compact horizonless objects, collectively referred to as black hole mimickers, have been proposed as alternatives to black holes. The main motivation is theoretical, to sidestep thorny issues connected with curvature singularities and event horizons in classical and semiclassical gravity. The study of black hole mimickers is also driven by astrophysical observations, to identify potential signatures that can distinguish non-singular compact objects from true black holes. See \cite{Bambi:2025wjx} for a recent status report of the field. 

In the present paper, we focus on so-called black shells, which are a particular form of black hole mimickers motivated by string theory \cite{Danielsson:2017riq}. The main assumption of the black shell proposal is that the vacuum is unstable against the formation of a bubble of AdS spacetime, contained within a dynamical shell formed by string theory 0-branes dissolved in a 2-brane. Open strings ending on the branes, carry a large number of degrees of freedom with entropy comparable to that of a black hole of the same mass.

The area of a static black shell is determined by the Israel-Lanczos junction conditions \cite{Lanczos:1924bgi,Israel:1966rt}, supplemented by the condition that the matter on the shell, separating the exterior geometry from the interior empty AdS spacetime, should have the equation of state of massless radiation \cite{Danielsson:2017riq}. For a Schwarzschild exterior, the shell is located at the so-called Buchdahl radius, $9/8$ times the Schwarzschild radius. The matching procedure is easily adapted to static black shells with electric charge and a Reissner-Nordstr\"om exterior \cite{Danielsson:2017riq}, while spinning black shells have been considered in \cite{Danielsson:2021ruf}.

If black shells are to succeed as black hole mimickers, replacing black holes in Nature, then they have to be the natural endpoint of gravitational collapse rather than black holes. In this case, when matter is on the verge of collapse into a black hole, a bubble of the true vacuum nucleates and the collapsing matter is converted into a black shell with an area that exceeds that of the would-be event horizon. {\it A priori}, such a nucleation, which is non-local and involves quantum tunneling on a macroscopic scale, may seem unlikely but the enormous available phase space associated with the large number of degrees of freedom of the black shell makes it more plausible. A similar argument for fuzzball formation is discussed in \cite{Mathur:2024ify}. 

Without the right dynamical ingredients, such a black shell is unstable and either expands, which would spell disaster for the universe, or collapses into a black hole. As shown in \cite{Danielsson:2021ykm,Giri:2024cks}, it is possible to stabilize the shell against small radial perturbations by introducing a mechanism that transfers energy between the tension of the 2-brane the radiation carried by the brane. A detailed string-theoretic derivation of such a mechanism is still lacking but see \cite{Danielsson:2021ruf} for suggestive speculations.

In the following, we consider the  thermodynamics of black shells in the canonical ensemble at fixed temperature and compare to black hole thermodynamics.
As long as our goal is restricted to equilibrium thermodynamics, we do not need to address the issue of dynamical black shell formation in gravitational collapse, nor argue for or against black shells replacing black holes.
In thermal equilibrium, the existence of black shell configurations, independent of any hypothetical stabilization mechanism, is sufficient in order to compare to black holes.

We work in Euclidean signature throughout and embed the system into an external AdS-spacetime in order to have a well-defined canonical ensemble. A small negative cosmological constant can be viewed as a regulator if the system of interest is a black shell in asymptotically flat spacetime. Alternatively, an asymptotically AdS black shell is a candidate gravitational dual to a thermal state in a strongly coupled boundary theory. 

\begin{figure}
  \centering
  \includegraphics[width=0.65\linewidth]{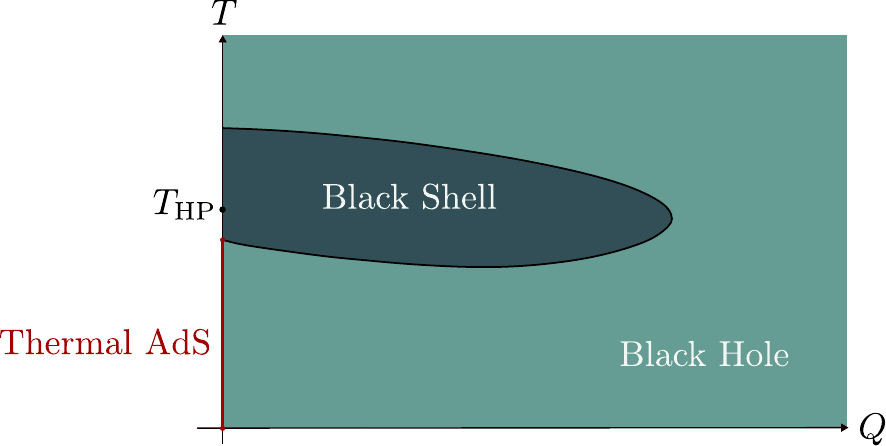}
  \caption{\small The plot indicates the configuration with the lowest free energy for any given values of temperature and charge. The red line segment along $Q=0$ corresponds to the thermal AdS solution. $T_{\text{HP}}$ denotes the Hawking-Page transition temperature.}
  \label{TvsQfigure}
  \end{figure}
  
Our results are summarized in Figure~\ref{TvsQfigure}. We find a rich structure, with phase transitions between black holes, black shells and the thermal vacuum in the canonical ensemble.
In Section~\ref{introsec}, we obtain the free energy of an electrically neutral black shell. We find that the well-known Hawking-Page phase transition between thermal AdS spacetime and AdS-Schwarzschild black holes~\cite{Hawking:1982dh} is split in two. There is a phase transition from thermal AdS spacetime to a phase dominated by black shells at a temperature of order the AdS scale but below the Hawking-Page temperature. Then there is a second phase transition from the black-shell phase to a more conventional black-hole phase at a temperature somewhat higher than the Hawking-Page temperature.

The situation is more involved when we allow the black shells to carry electric charge. Under gauge/gravity duality, this amounts to turning on a conserved global $U(1)$ charge in the boundary gauge theory. At fixed non-vanishing charge, below a certain critical value, there are phase transitions from black holes to black shells, and back again to black holes, as the temperature is increased. Above the critical charge, however, black holes dominate the ensemble at all temperatures.

The remainder of the paper is organized as follows. We begin in Section~\ref{introsec} with a brief review of the original black shell construction of \cite{Danielsson:2017riq}. We then obtain a closed form expression for the free energy of an electrically neutral black shell that can be compared to the free energy of an AdS-Schwarzschild black hole. In Section~\ref{finiteQ} we repeat the matching procedure for black shells with non-vanishing electric charge and numerically evaluate the resulting free energy as a function of charge and temperature in the canonical ensemble. In Section~\ref{interpreting} we conclude with a brief discussion of our results and their possible relevance to the suggested role of black shells as black hole mimickers.

\section{Black Shells in Asymptotically AdS$_{3+1}$ Spacetime}
\label{introsec}
We begin with a brief review of the black shell construction of \cite{Danielsson:2017riq} with the added twist of a small negative cosmological constant outside the shell in order to have a finite free energy in the canonical ensemble. In this case, the spacetime is asymptotically AdS rather than asymptotically flat as in \cite{Danielsson:2017riq}. 

Consider a spherically symmetric shell of radius $r_0$ in 3+1 spacetime dimensions. Inside the shell, the spacetime is taken to be AdS with a characteristic length scale $\ell$, while outside the shell, it is described by an AdS-Schwarzschild geometry with mass $M$ and AdS length scale $L\gg\ell$. The metric is then given by:
\bea
\begin{aligned}
\dd{s}_{-}^2 &= -\left(\frac{r^2}{\ell^2}+1\right)\dd t_{-}^2 + \frac{\dd r^2}{\left(\frac{r^2}{\ell^2}+1\right)} + r^2 \dd \Omega_2^2\, \quad \quad &\text{if} \ r<r_0\\
\dd{s}_{+}^2 &= -\left(\frac{r^2}{L^2}+1-\frac{2M}{r}\right)\dd t_{+}^2 + \frac{\dd r^2}{\left(\frac{r^2}{L^2}+1-\frac{2M}{r}\right)} + r^2 \dd \Omega_2^2\, \quad \quad &\text{if} \ r>r_0
\end{aligned}
\eea 
We work in units where Newton's constant $G_N$ is set to 1 and denote the time coordinates outside and inside the shell by $t_+$ and $t_-$, respectively. We use the area of the two-transverse sphere to label the radial position, via $A=4\pi r^2$, both outside and inside the shell.

The parameters of the metric are constrained by the usual junction conditions \cite{Israel:1966rt,Lanczos:1924bgi,Sen:1924ons}:
\begin{itemize}
  \item The induced metric is continuous across the shell 
  \bea
  h_{ij}^{+} = h_{ij}^{-}\,.
  \label{firstjunctioncondition}
  \eea 
  \item The discontinuity in the extrinsic curvature is related to the stress-energy tensor on the shell:
  \bea
  K_{i j}^{+}-K_{i j}^{-}=-8 \pi \left(S_{i j}-\frac{1}{2} h_{i j} h^{k l} S_{k l}\right)\,. \label{secondjunctioncondition}
  \eea
\end{itemize}
Using the definition of the induced metric,
\bea
h_{ij} = g_{kl}\frac{\partial X^{k}}{\partial \sigma^i}\frac{\partial X^l}{\partial \sigma^j}\,,
\eea 
and identifying the time coordinate $\sigma^0$ on the shell with the exterior AdS-Schwarzschild time  $t_+$, the first junction condition relates the time coordinates across the shell,
\bea
\frac{\dd t_{-}}{\dd t_{+}} = \sqrt{\frac{\frac{r_0^2}{L^2}+1-\frac{2M}{r_0}}{\frac{r_0^2}{\ell^2}+1}}\,.
\eea 
To proceed further, we take the stress-energy tensor on the shell to be that of a perfect fluid at rest,
\bea
\label{shellstress}
S_{i j}=(\rho+p) u_i u_j+p h_{i j}\,,
\eea 
where $\rho$ and $p$ are, respectively, the energy density and pressure of the fluid, and the components of the fluid 3-velocity, 
\bea
u^{t_{+}} = \frac{1}{\sqrt{\frac{r_0^2}{L^2}+1-\frac{2M}{r_0}}}\,, \quad  \ u^{\phi}=u^{\theta}=0\,.
\eea
satisfy $u\cdot u=-1$. 
This gives us 
\bea
S_{t t}=\rho\left(\frac{r_0^2}{L^2}+1-\frac{2 m}{r_0}\right)\,, \quad S_{\theta \theta}=p\, r_0^2\,, \quad S_{\phi \phi}=p\, r_0^2 \sin ^2 \theta\,, \quad h^{k l} S_{k l}=-\rho+2 p\,.
\eea 
The second junction condition \eqref{secondjunctioncondition} relates the shell pressure and energy density. For a metric of the form 
\bea
\dd s^2  = N^2(r) \dd{r}^2 + h_{ij}\dd x^{i}\dd x^{j}\,,
\eea 
the extrinsic curvature $K_{ij}$ of a constant $r$ surface is given by $\frac{1}{2N}\frac{\partial h_{ij}}{\partial r}$. This gives us
\begin{alignat}{2}
K_{t t}^{+}&=-\left(\frac{m}{r_0^2}+\frac{r_0}{L^2}\right) \sqrt{\frac{r_0^2}{L^2}+1-\frac{2 m}{r_0}}\,, \quad \quad   && K_{\theta \theta}^{+}=r_0 \sqrt{\frac{r_0^2}{L^2}+1-\frac{2 m}{r_0}}\,, \\
K_{t t}^{-}&=-\frac{r_0}{\ell^2}\left(\frac{\frac{r_0^2}{L^2}+1-\frac{2 m}{r_0}}{ \sqrt{\frac{r_0^2}{\ell^2}+1}}\right), && K_{\theta \theta}^{-}=r_0 \sqrt{\frac{r_0^2}{\ell^2}+1}\,.
\end{alignat}
We now plug these expressions into the junction condition \eqref{secondjunctioncondition}, and solve for the energy density and pressure,
\bea
\rho=\frac{1}{4 \pi r_0}\left(\sqrt{\frac{r_0^2}{\ell^2}+1}-\sqrt{\frac{r^2_0}{L^2}+1-\frac{2 M}{r_0}}\right)\,, \label{rhoofreq}
\eea  
\bea
p =\frac{1}{8 \pi r_0}\left(\frac{\frac{2 r^2_0}{L^2}+1-\frac{M}{r_0}}{\sqrt{\frac{r^2_0}{L^2}+1-\frac{2 M}{r_0}}}-\frac{1+\frac{2r_0^2}{\ell^2}}{\sqrt{\frac{r_0^2}{\ell^2}+1}}\right)\,.
\eea
Formula \eqref{rhoofreq} for the energy density can be rearranged as follows,
\bea
4 \pi r_0^2  \rho - r_0 \left(\sqrt{\frac{r_0^2}{\ell^2}+1}-\sqrt{1+\frac{r_0^2}{L^2}}\right) = r_0 \left(\sqrt{\frac{r_0^2}{L^2}+1} - \sqrt{\frac{r_0^2}{L^2}+1-\frac{2 M}{r_0}}\right) \,.\label{rearranged} 
\eea 
The terms on the left hand side can be interpreted as the energy of the shell and a negative energy of the AdS bubble inside the shell. These contributions add up to the total energy of the AdS-Schwarzschild geometry (with respect to empty AdS spacetime) on the right,
\bea
E = r_0 \left(\sqrt{\frac{r_0^2}{L^2}+1} - \sqrt{\frac{r_0^2}{L^2}+1-\frac{2 M}{r_0}}\right) \,.\label{energyequation}
\eea 
The total energy of the shell configuration includes gravitational self-interaction and satisfies the relation,
\bea
M=E\sqrt{\frac{r_0^2}{L^2}+1}-\frac{E^2}{2r_0}\,,
\eea 
from which it is easily checked that $0<E<M$ for any $M>0$.

The energy density of a black shell of energy E and radius $r_0$ is then
\bea
\rho_b\equiv \frac{E}{4\pi r_0^2}
=\frac{1}{4 \pi r_0}\left( \sqrt{\frac{r_0^2}{L^2}+1} - \sqrt{\frac{r_0^2}{L^2}+1-\frac{2 M}{r_0}}\right) \,.
\label{energydensity}
\eea
For a dynamical shell of radius $r$, the surface energy density $\rho_b(r)$ combines with a surface pressure $p_b(r)$ in a shell stress-energy tensor of the form \eqref{shellstress}. The relation
\bea
p_b=-\rho_b-\left.\frac{r}{2} \frac{d \rho_b}{d r}\right|_{r=r_0} \,
\label{energyconservationeq}
\eea
follows from the conservation of the stress-energy tensor on the shell. Substituting the expression \eqref{energydensity} for $\rho_b$ then leads to the following expression for the surface pressure,
\bea
p_b = \frac{1}{8 \pi r_0}\left(\frac{\frac{2 r^2_0}{L^2}+1-\frac{M}{r_0}}{\sqrt{\frac{r^2_0}{L^2}+1-\frac{2 M}{r_0}}}-\frac{1+ \frac{2 r^2_0}{L^2}}{\sqrt{1+\frac{r^2_0}{L^2}}}\right)\,.
\eea 
Furthermore, if we take the equation of state on the shell to be that of radiation,
\bea
\label{eos}
\rho_b =2 p_b\,,
\eea
then we find that the radius of the shell is given by the simple expression
\bea
M = \frac{4 r_0}{9} \left(\frac{1+\frac{3r^2_0}{2L^2}}{1+\frac{r^2_0}{L^2}}\right)\label{Mofreq}
\eea
When $L\gg r_0$, we reproduce the Buchdahl radius in \cite{Danielsson:2017riq}. When $L \ll r$, we find that $r_0 = \frac32 M$.

We note that the radiation equation of state in \eqref{eos} is not derived from first principles, but it is motivated by string theory, by viewing the shell as a collection of D0-particles dissolved in a spherical D2-brane, with a gas of massless open strings ending on the D-branes \cite{Danielsson:2017riq}. Having dynamical matter on the shell is crucial to its stability against small perturbations and also provides a mechanism for exchanging energy with a thermal environment.

\begin{figure}
  \centering
  \includegraphics[width=0.5\linewidth]{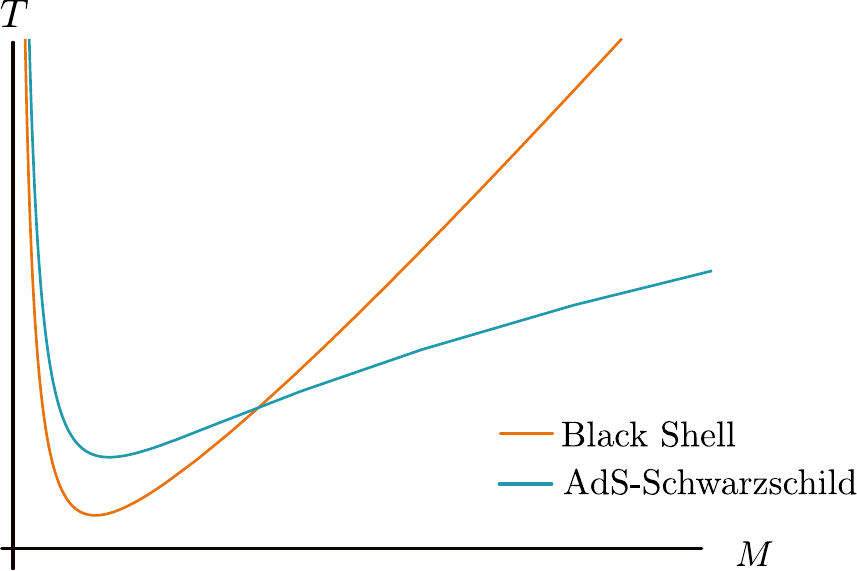}
  \caption{\small The temperature of black holes and black shells versus their mass.}
  \label{tvsMfig}
  \end{figure}

We are interested in the thermodynamics of these black shell configurations. Let us start by calculating the temperature of the system. The gas of open strings on the shell is not in free fall and accordingly it will equilibrate at the local Unruh temperature \cite{Saravani:2012is,Danielsson:2017riq}. If $a$ is the proper acceleration of an observer sitting at a constant $r_0$ surface, the corresponding local Unruh temperature is given by
\bea
T_U=\frac{a}{2 \pi}=\frac{1}{2 \pi r_0^2} \frac{M+\frac{r_0^3}{L^2}}{\sqrt{\frac{r_0^2}{L^2}+1-\frac{2 M}{r_0}}}\,.
\eea
Stripping off the redshift factor, we obtain the following shell temperature
\bea
T_S = \frac{1}{2 \pi r_0^2} \left(M+\frac{r_0^3}{L^2}\right)\,,\label{Tofreq}
\eea 
shown in Figure~\ref{tvsMfig}, where we have included the Hawking temperature of an AdS-Schwarzschild black hole of the same mass, for comparison. We note that in both cases there is a minimum temperature, of order the AdS length scale, that separates two branches, {\it i.e.} small and large black objects. Small black holes and small black shells have negative specific heat while their large counterparts have positive specific heat and can therefore be thermodynamically stable. In remainder of this section, we investigate the question of thermodynamic stability by evaluating and comparing the free energy of black holes and black holes at fixed temperature.

The entropy can be obtained via the thermodynamic relation 
\bea
\frac{\dd S}{ \dd M} = \frac{1}{T}\,.  \label{firstlawthermo}
\eea 
For an AdS-Schwarzschild black hole we have
\bea
T_{BH} =  \frac{1}{4 \pi r_h}\left(1+\frac{3r_h^2}{L^2}\right)\,,\qquad
M_{BH}=\frac{r_h}{2}\left(\frac{r_h^2}{L^2}+1\right),
\eea 
which gives expected result,
\bea
S_{BH}=\int_0^{r_h}\dd r\, \frac{1}{T_{BH}}\frac{\dd M_{BH}}{\dd r}=\pi r_h^2\,.
\eea 
For a black shell, we instead use \eqref{Mofreq} and \eqref{Tofreq} for the mass and temperature, respectively, and obtain 
\bea
S = \frac{2}{3} \pi  L^2 \left(3 \log \left(L^2+r^2_0\right)-2 \log \left(L^3+\frac{3 L r^2_0}{4}\right)\right)\label{Sofreq}\,.
\eea 
When $L \gg r$, the leading contribution to the entropy is the standard area term
\bea
S \simeq \pi r_0^2 + O\left(\frac{r_0^4}{L^2}\right)\,.
\eea 
The black shell free energy is then given by
the following expression, 
\bea
\begin{aligned}
F &=M- T_S S\,,\\
&= \frac{2 r \left(2 L^2+3 r_0^2\right)}{9 \left(L^2+r_0^2\right)}-\frac{15 L^2 r_0^2+4 L^4+9 r_0^4}{27  r^3+27 L^2 r_0} \left[3 \log \left(L^2+r^2_0\right)-2 \log \left(L^3+\frac{3 L r^2_0}{4}\right)\right]\,.
\end{aligned}
\eea 

\begin{figure}
  \centering
  \includegraphics[width=0.5\linewidth]{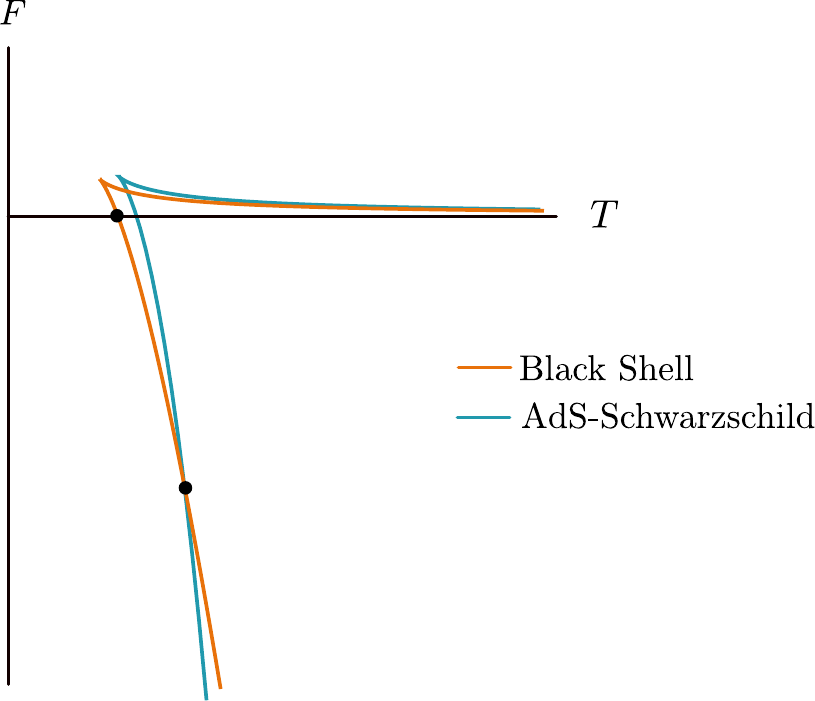}
  \caption{\small Free energies of black holes and black shells as functions of temperature. Thermal AdS spacetime has zero free energy and therefore corresponds to the horizontal axis. The two first order phase transitions are marked by black dots.}
  \label{freeenergyfig}
  \end{figure}
  
In Figure~\ref{freeenergyfig}, we plot the free energy as a function of temperature, and for comparison, we also include the free energy of an AdS-Schwarzschild black hole, which is given by
\bea
F_{BH} = \frac{r_h}{4}\left(1-\frac{r_h^2}{L^2}\right)\,.
\eea 
The plot reveals interesting phase structure. In the absence of black shells the canonical ensemble is dominated by thermal AdS spacetime (with $F=0$) at low temperature and undergoes the well-known Hawking-Page transition to a phase dominated by large black holes at a temperature of order the AdS length scale. If the black shell solutions considered here are indeed allowed configurations in the thermal ensamble then the Hawking-Page transition to black holes is preempted by a first-order transition to a black-shell phase at a temperature below the Hawking-Page temperature (but still of order the AdS scale). The system then has another first-order phase transition to a large black hole phase at a temperature higher than the usual Hawking-Page temperature. The two phase transitions are indicated by black dots in Figure~\ref{freeenergyfig}.

The existence of black shells has immediate implications for holographic duality at finite volume in that there is a range of temperatures for which the gravitational dual description of a thermal state is given by a black shell rather than a black hole. This curious fact merits further study but it does not appear to require any major revision of existing results. This is because most applications of gauge/gravity duality to strongly coupled quantum critical systems involve taking a planar limit on the gravitational side of the duality, for which planar AdS black holes still dominate the ensemble.

\section{Electrically Charged Black Shells}
\label{finiteQ}

It is straightforward to generalize our considerations to black shells carrying electric charge. 
As in the previous section, we will denote the radius of the shell by $r_0$. Inside the shell, the spacetime is AdS with a length scale $\ell$, while outside the shell, we have an AdS-RN black hole with AdS length scale $L\gg\ell$. This gives us the metric:
\bea
\begin{aligned}
\dd{s}_{-}^2 &= -\left(\frac{r^2}{\ell^2}+1\right)\dd t_{-}^2 + \frac{\dd r^2}{\left(\frac{r^2}{\ell^2}+1\right)} + r^2 \dd \Omega_2^2\, \quad \quad &\text{when} \ r<r_0\\
\dd{s}_{+}^2 &= -f(r)\dd t_{+}^2 + \frac{\dd r^2}{f(r)} + r^2 \dd \Omega_2^2\, \quad \quad &\text{when} \ r<r_0
\end{aligned}
\eea 
where 
\bea
f(r) = \frac{r^2}{L^2}+1-\frac{2M}{r}+\frac{Q^2}{r^2}\,.
\eea 
Using the junction conditions \eqref{firstjunctioncondition} and \eqref{secondjunctioncondition}, we find that
\bea
\rho=\frac{1}{4 \pi r_0}\left(\sqrt{\frac{r^2_0}{\ell^2}+1}-\sqrt{\frac{r^2_0}{L^2}+1-\frac{2 M}{r_0}+\frac{Q^2}{r^2}}\right)\,, \label{rhoofrcharged}
\eea 
and 
\bea
p =\frac{1}{8 \pi r_0}\left(\frac{\frac{2 r^2_0}{L^2}+1-\frac{M}{r_0}}{\sqrt{\frac{r^2_0}{L^2}+1-\frac{2 M}{r_0}+\frac{Q^2}{r^2}}}-\frac{1+\frac{2r_0^2}{\ell^2}}{\sqrt{\frac{r_0^2}{\ell^2}+1}}\right)\,.
\eea
Once again, we can rewrite the junction condition involving $\rho$ as in \eqref{rearranged}. This gives us the following expression for the total energy,
\bea
E = r_0 \sqrt{1+\frac{r_0^2}{L^2}} - r_0 \sqrt{\frac{r_0^2}{L^2}+1-\frac{2 M}{r_0}+\frac{Q^2}{r^2}} \equiv 4\pi r_0^2 \rho_b\,.
\eea 
Using the conservation equation  \eqref{energyconservationeq}, we find that the pressure on the shell is given by
\bea
p_b = \frac{1}{8 \pi r_0}\left(\frac{\frac{2 r^2_0}{L^2}+1-\frac{M}{r_0}}{\sqrt{\frac{r^2_0}{L^2}+1-\frac{2 M}{r_0}+\frac{Q^2}{r^2}}}-\frac{1+ \frac{2 r^2_0}{L^2}}{\sqrt{1+\frac{r^2_0}{L^2}}}\right)\,.
\eea 
As before, we impose a radiation equation of state $\rho_b =2 p_b$ on the shell. Then the shell radius is related to the mass $M$ via the relation
\bea
M=\frac{1}{9}\frac{(3r_0^2+2L^2)}{(r_0^2+L^2)}\left(r_0+\sqrt{r_0^2+\frac{3Q^2}{L^2}\left(r_0^2+ L^2\right)}\,\right)+\frac{Q^2}{3r_0},
\eea
giving a family of charged black shell solutions that reduce to the uncharged case, considered above, in the $Q\rightarrow 0$ limit.

\begin{figure}
  \centering
  \hspace{-0.4cm}\includegraphics[width=1.01\linewidth]{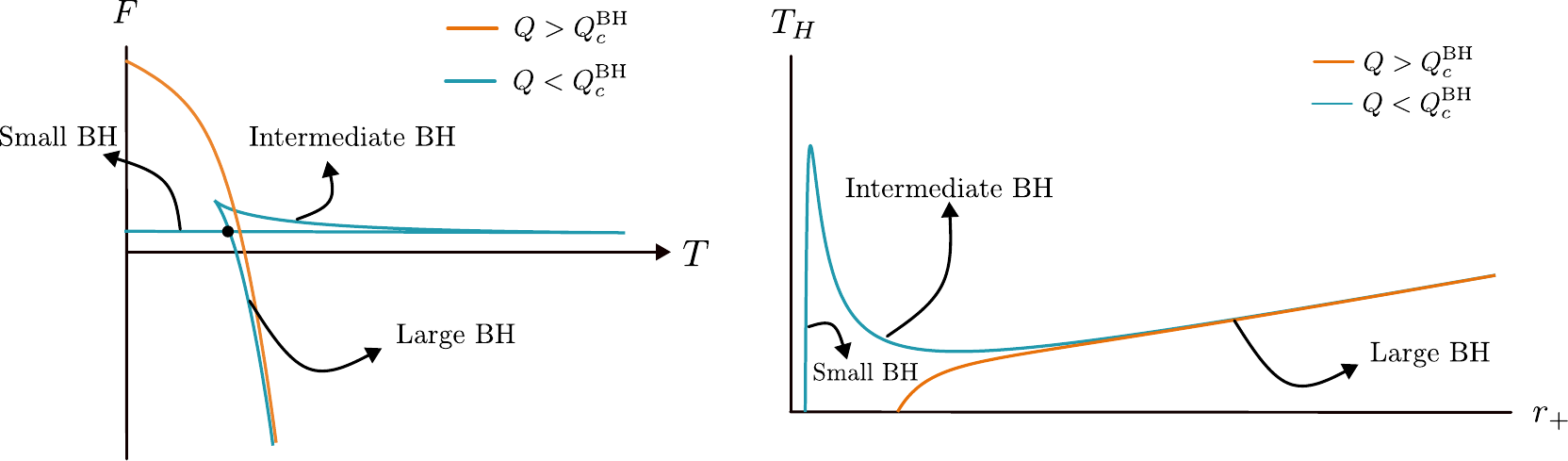}
  \caption{\small \textit{(Left)} Free energy of a charged black hole as a function of temperature for two different values of the charge. \textit{(Right)} Temperature of the black hole as a function of the event horizon radius for the same charge values. In both plots, three branches of black holes are visible below the critical charge $Q<Q_c^{\text{BH}}$.}
  \label{chargedbhfigure}
  \end{figure}

Before computing the free energy of the black shell, it is instructive to study the black hole case first. Working in the fixed charge ensemble, we find that 
\bea
F_{BH} = -\frac{r_+^3}{4 L^2}+\frac{3 Q^2}{4 r_+}+\frac{r_+}{4}\,, \quad \quad T_{BH} =  \frac{1}{4 \pi r_+}\left(\frac{3 r_+^2}{L^2}-\frac{Q^2}{r_+^2}+1\right)\,.
\eea 
Let us first examine the behaviour of the black hole temperature $T_{BH}$. Figure~\ref{chargedbhfigure} plots the temperature as a function of $r_+$ for two different charge values. The blue curve corresponds to a charge below a critical value, denoted by $Q_c^{\text{BH}}$. For a range of temperatures, a line of constant $T$ will intersect the plot at three different values of $r_+$. These branches correspond to small, intermediate, and large black holes. For black hole charge above the critical value, $Q>Q_c^{\text{BH}}$, only a single branch remains. For $Q\gg Q_c^{\text{BH}}$, this branch corresponds to the large black hole solutions.

Next, let us look at the free energy of the black hole as a function of temperature. When $Q<Q_c^{\text{BH}}$, we find classic swallowtail behaviour. Each segment of the swallowtail corresponds to the branches we identified in the temperature plot (see Figure \ref{chargedbhfigure}). Only small black holes exist at low temperature. As we increase $T$, we find a range of temperatures where all three black hole branches coexist. Inside this range there is a critical temperature where a first-order phase transition (indicated by a black dot in Figure~\ref{chargedbhfigure}) occurs from the low-temperature small black hole phase to a high-temperature large black hole phase. When $Q>Q_c^{\text{BH}}$, there is no phase transition as there is only a single branch. 

The value of the critical charge $Q_{c}^{\text{BH}}$ is easily determined by requiring that the local maximum and local minimum of $T_{BH}$ merge at a single value of $r_+$, giving
\bea
Q_c^{\text{BH}} = \frac{L}{6}\,.\label{BHcriticalcharge}
\eea

  \begin{figure}
    \centering
    \includegraphics[width=0.6\linewidth]{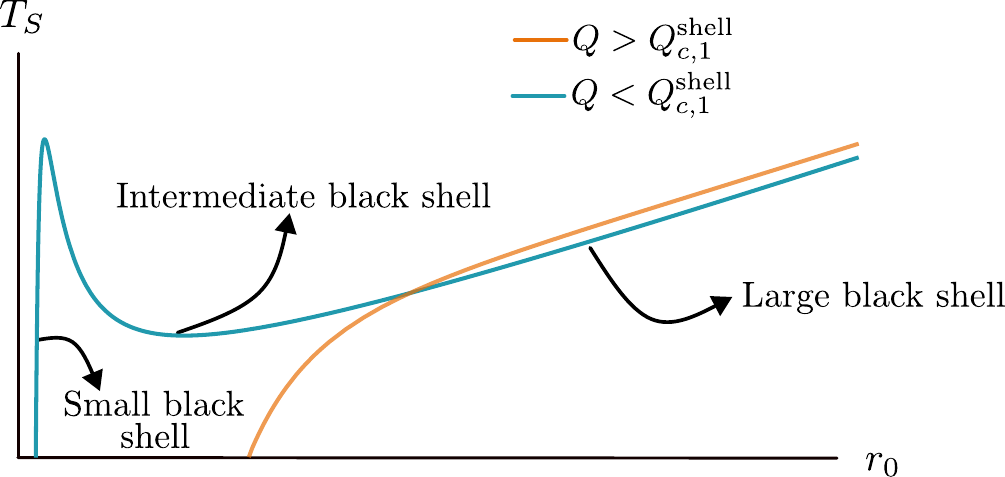}
    \caption{\small Temperature of a black shell as function of its radius. When $Q<Q_{c,1}^{\text{shell}}$, three branches appear, corresponding to small, intermediate, and large black shells. When $Q>Q_{c,1}^{\text{shell}}$, the branches disappear.}
    \label{shelltempfigure} 
    \end{figure}
    
Now, we turn our attention to the charged black shell configuration and compute its temperature and free energy. The proper acceleration in the rest frame of the shell surface at $r=r_0$ gives the shell temperature,
\bea
T_S = \frac{1}{2 \pi r_0^2} \left(M+\frac{r_0^3}{L^2}-\frac{Q^2}{r_0}\right)\,.
\eea 
As in the black hole case, there is a critical value of the charge, $Q=Q_{c,1}^{\text{shell}}$, below which there are three branches which we refer to as small, intermediate, and large black shells (see Figure~\ref{shelltempfigure}). Above the critical charge, $Q>Q_{c,1}^{\text{shell}}$, there is only a single branch. We find that $Q_{c,1}^{\text{shell}}$ is always larger than the corresponding black hole critical charge, $Q_{c}^{\text{BH}}$.

\begin{figure}
  \centering
  \includegraphics[width=1\linewidth]{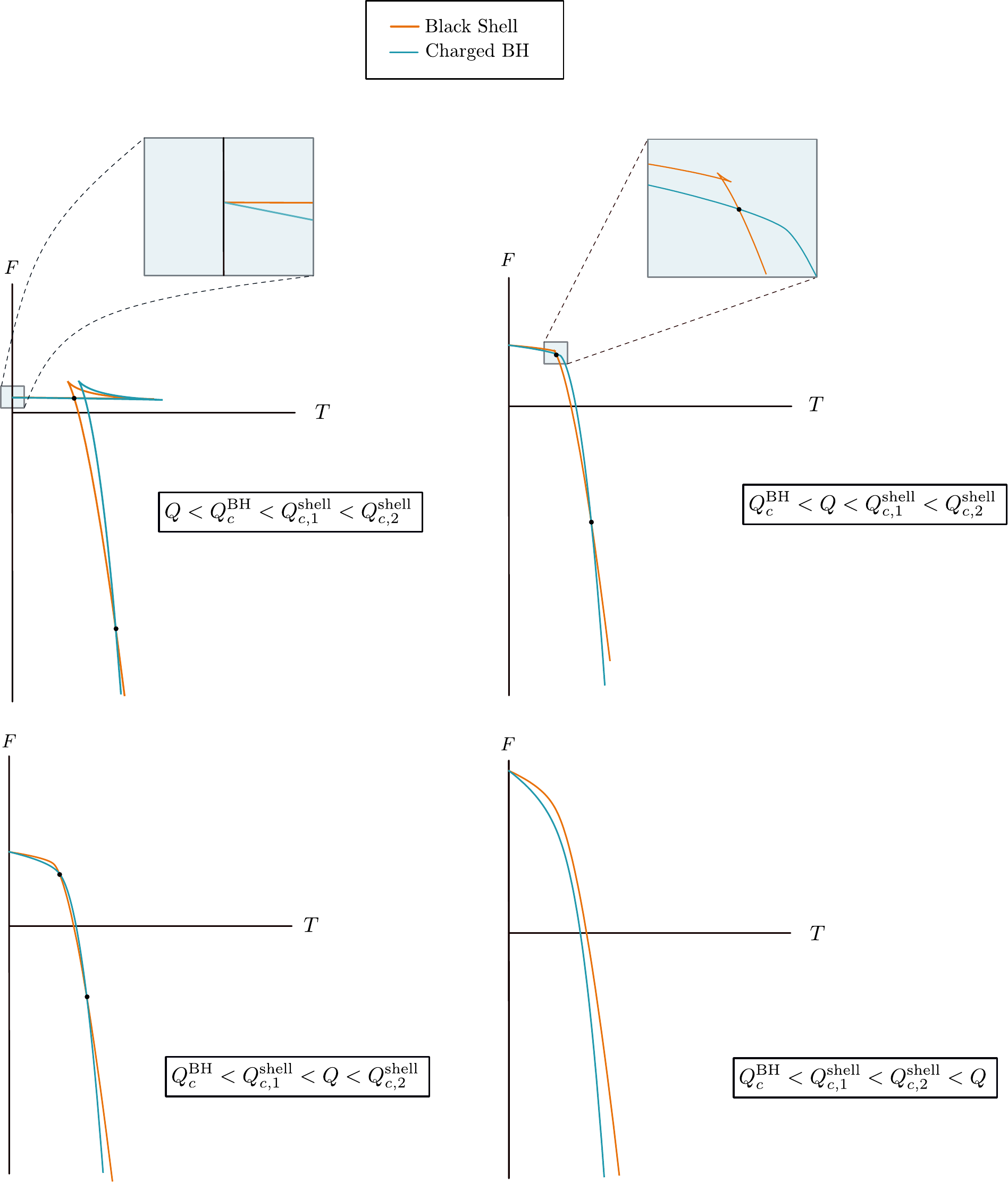}
  \caption{\small Comparison of free energies of black shells and black holes at various charges. First-order phase transitions are indicated by black dots.}
  \label{chargedshellfigure}
  \end{figure}

Since we work in a fixed charge ensemble, the first law, $\dd{M} = T \dd{S}+ \mu \dd{Q}$, reduces to \eqref{firstlawthermo}. This allows us to find $S$ as a function of $r_0$. We can also compute the free energy using the relation $F=M-T S$. Unfortunately, closed-form expressions are not available for these quantities, but they are easily evaluated numerically. In Figure~\ref{chargedshellfigure} we plot the resulting free energy as a function of the shell temperature for various values of the charge. In each case, the free energy of a black hole with the same charge is included for comparison.

Let us analyze these plots in some detail. 
First of all, in the limit of zero temperature, black holes and black shells have the same free energy and can therefore coexist. This holds for any non-vanishing value of the charge. The zero temperature limit corresponds to the extremal limit of the charged black hole. By comparing free energies at finite temperature, we uncover a rich phase structure that depends on the value of the electric charge carried by the compact object. 

At small charge, $Q<Q_c^{\text{BH}}<Q_{c,1}^{\text{shell}}$, the free energies of black shells and black holes both display swallowtail behaviour, where each segment corresponds to the small, intermediate or large branch of black shells and black holes, respectively. At low temperature, the black hole free energy is lower than that of black shells, making small black holes  thermodynamically favored in this temperature regime. As we increase the temperature we reach a critical point where the system undergoes a first order phase transition to a phase dominated by large black shells. This phase persists until we reach a second critical temperature, where there is another first order phase transition to a large black hole phase.
For all temperatures above this point, large black holes dominate with the lowest free energy.

Next, let us consider a low-to-intermediate charge case, where \( Q_c^{\text{BH}} < Q < Q_{c,1}^{\text{shell}} \). We observe the same phase transition structure as in the previous case. The only difference is that the black hole free energy no longer exhibits swallowtail behaviour.

Finally, we consider the case where the charge is above both critical values, \( Q_c^{\text{BH}} < Q_{c,1}^{\text{shell}} < Q \) and the swallowtail structure in the black shell free energy plot has also disappeared. In this regime, a new scale emerges, which we denote by \( Q_{c,2}^{\text{shell}} \). For \( Q_c^{\text{BH}} < Q_c^{\text{shell}} < Q < Q_{c,2}^{\text{shell}} \), the same phase transition structure persists. However, when \( Q > Q_{c,2}^{\text{shell}} \), the phase transitions disappear and black holes always have the lowest free energy. The only temperature at which black holes and black shells can co-exist in this regime is at the extremal limit of the black hole.

\section{Interpreting the results}
\label{interpreting}

Black shells have been proposed as black hole mimickers that replace black holes. An important motivation is to provide a resolution to the black hole information paradox. For that to work, black shells have to be overwhelmingly favored over black holes as the endpoint of gravitational collapse of matter. We do not address this crucial dynamical question here, but our results on the equilibrium thermodynamics of black shells versus black holes can be viewed as indirect evidence for black shells being favored in a certain range of parameters.  
  
\begin{figure}[h!]
  \centering
  \includegraphics[width=0.6\linewidth]{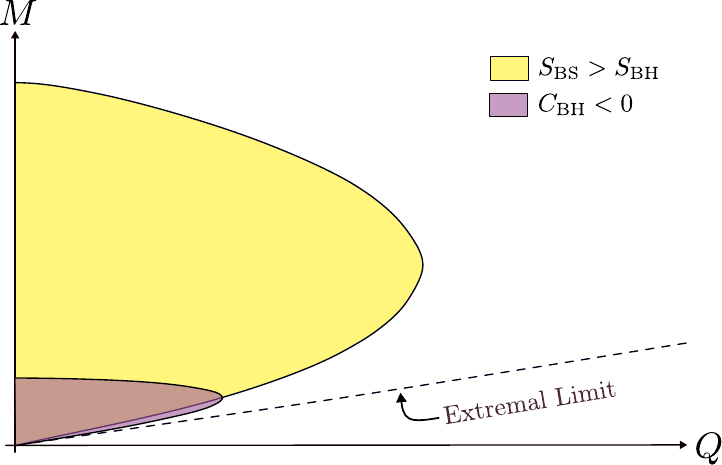}
  \caption{\small The region shaded in yellow corresponds to the values of $M$ and $Q$ for which the black shell entropy is greater than the black hole entropy at a given mass. The purple region indicates where black holes have negative specific heat $C_{\text{BH}}$.}
  \label{MvsQfigure2}
  \end{figure}

For the purposes of this paper, we have assumed the possible coexistence of black shells and black holes, and investigated thermodynamical transitions between the two types of objects. We have uncovered a rather rich structure in the phase diagram. For instance, as illustrated in Figure~\ref{TvsQfigure}, we find that the Hawking-Page transition between the thermal AdS vacuum and neutral AdS-Schwarzschild black holes is split across a black shell phase. For non-vanishing charge, below a certain critical value, the phase transition between small and large black holes is similarly split, revealing an intermediate black shell phase. 

If we want to say something about the information paradox, the most interesting black holes are those with negative specific heat, which evaporate in the absence of a heat bath and potentially lead to a paradox. In Figure~\ref{MvsQfigure2} we highlight in purple the region where black holes have negative specific heat. In the same diagram, we also indicate in yellow the values of $M$ and $Q$ for which the entropy of a black shell is higher than for the corresponding black hole. According to the black shell proposal any black hole that threatens to form in this region, would instead transit into a black shell of the same mass due to its higher entropy. We observe that all AdS-Schwarzschild black holes with negative specific heat are inside the yellow region  The same is true for most charged black holes with negative specific heat as well. 
Still, there is a narrow strip of purple  outside of the yellow region in Figure~\ref{MvsQfigure2}, that contains charged black holes with a negative specific heat where the black hole entropy is higher than the one of a black shell. However, in view of the weak gravity conjecture, such charged black holes are expected to discharge themselves through emission of particles with more charge than mass. This brings the system into the yellow region in Figure~\ref{MvsQfigure2}, where, according to the black shell proposal, the black hole can safely nucleate into a non-singular black shell before continuing to evaporate. 

Our results on black shell thermodynamics thus appear to be consistent with the black shell proposal of \cite{Danielsson:2017riq} but they do not offer any direct insight into the key dynamical questions concerning the production and stability of black shells. Gaining a better theoretical handle on quantum tunneling into high-entropy final states is an important goal, with implications for a wide range of theoretical problems, and not limited to the area of black hole mimickers. 

More specific to black shells, is the issue of their stability against small radial perturbations. While the considerations in  \cite{Danielsson:2021ykm,Giri:2024cks} are suggestive, a top-down derivation from string theory of a mechanism to support their stability would settle the matter.
Subject to this caveat, black shells appear to be valid solutions of string theory and as such they feature in any thermodynamic ensembles involving gravity. 
Regardless of whether they can serve as black hole mimickers, we find our results on black shell thermodynamics intriguing.

\section*{Acknowledgements} 
We would like to thank Diego Hidalgo for useful
discussions. This work was supported in part by the Icelandic Research Fund under grant 228952-053. VM is supported by a doctoral grant from The University of Iceland Science Park. Support from Kungliga Vetenskapssamhället i Uppsala and Kungliga Fysiografiska sällskapet i Lund is also acknowledged.

\bibliographystyle{JHEP}
\bibliography{refs}

\end{document}